\documentclass[longauth, bibyear]{aa}
\usepackage{graphicx}
\usepackage{txfonts}
\usepackage[colorlinks=true, 
urlcolor=blue, 
citecolor=blue, 
linkcolor=blue]{hyperref}
\usepackage{natbib}
\usepackage{booktabs}
\usepackage{xcolor}
\makeatletter
\renewcommand{\@fnsymbol}[1]{\textcolor{blue}{$\star$}}
\makeatother

\usepackage[normalem]{ulem}
\begin{document}

   \title{Constraining the TeV gamma-ray emission of SN 2024bch, a possible type IIn-L from a red supergiant progenitor }   
   \subtitle{Multiwavelength observations and analysis of the progenitor}
 \titlerunning{Constraining the TeV $\gamma$-ray emission of SN 2024bch, a possible type IIn-L from a RSG progenitor }  
    \author{
    K.~Abe\inst{1} \and
    S.~Abe\inst{2} \and
    A.~Abhishek\inst{3} \and
    F.~Acero\inst{4,5} \and
    A.~Aguasca-Cabot\thanks{Corresponding authors (alphabetical order): A. Aguasca-Cabot, A. Carosi, A. L\'opez-Oramas, A. Simongini; email: lst-contact@cta-observatory.org}\inst{,6} \and 
    I.~Agudo\inst{7} \and
    C.~Alispach\inst{8} \and
    D.~Ambrosino\inst{9} \and
    F.~Ambrosino\inst{10} \and
    L.~A.~Antonelli\inst{10}
% ...
    {(the CTAO-LST collaboration)}
    }
    \institute{
    Department of Physics, Tokai University, 4-1-1, Kita-Kaname, Hiratsuka, Kanagawa 259-1292, Japan
    \and Institute for Cosmic Ray Research, University of Tokyo, 5-1-5, Kashiwa-no-ha, Kashiwa, Chiba 277-8582, Japan
    \and INFN and Università degli Studi di Siena, Dipartimento di Scienze Fisiche, della Terra e dell'Ambiente (DSFTA), Sezione di Fisica, Via Roma 56, 53100 Siena, Italy
    \and Université Paris-Saclay, Université Paris Cité, CEA, CNRS, AIM, F-91191 Gif-sur-Yvette Cedex, France
    \and FSLAC IRL 2009, CNRS/IAC, La Laguna, Tenerife, Spain
    \and Departament de Física Quàntica i Astrofísica, Institut de Ciències del Cosmos, Universitat de Barcelona, IEEC-UB, Martí i Franquès, 1, 08028, Barcelona, Spain
    \and Instituto de Astrofísica de Andalucía-CSIC, Glorieta de la Astronomía s/n, 18008, Granada, Spain
    \and Department of Astronomy, University of Geneva, Chemin d'Ecogia 16, CH-1290 Versoix, Switzerland
    \and INFN Sezione di Napoli, Via Cintia, ed. G, 80126 Napoli, Italy
    \and INAF - Osservatorio Astronomico di Roma, Via di Frascati 33, 00040, Monteporzio Catone, Italy
% ...
}

    \date{Received 24 March, 2025; accepted 30 July, 2025}

    \abstract 
{
We present very high-energy optical photometry and spectroscopic observations of SN~2024bch in the nearby galaxy NGC~3206 ($\sim$ 20 Mpc). We used gamma-ray observations performed with the first Large-Sized Telescope (LST-1) of the Cherenkov Telescope Array Observatory (CTAO) and optical observations with the Liverpool Telescope (LT) combined with data from public repositories to evaluate the general properties of the event and the progenitor star. No significant emission above the LST-1 energy threshold for this observation ($\sim 100$~GeV) was detected in the direction of SN~2024bch, and we computed an integral upper limit on the photon flux of $F_\gamma(>100 \,\mbox{GeV}) \le 3.61 \times 10^{-12}$ cm$^{-2}$ s$^{-1}$ based on six nonconsecutive nights of observations with the LST-1, between 16 and 38 days after the explosion. Employing a general model for the gamma-ray flux emission, we found an upper limit on the mass-loss-rate to wind-velocity ratio of $\dot M/u_{\text{w}} \le 10^{-4} \frac{M_\odot}{\text{yr}}\frac{\text{s}}{\text{km}}$, although gamma-gamma absorption could potentially have skewed this estimation, effectively weakening our constraint. From spectro-photometric observations we found progenitor parameters of $\text{M}_{\text{pr}}=11$ -- $20$ $M_\odot$ and $\text{R}_{\text{pr}} = 531 \pm 125$ R$_\odot$. Finally, using archival images from the \textit{Hubble} Space Telescope, we constrained the luminosity of the progenitor star to log ($\text{L}_{\text{pr}}$/$L_\odot$) $\le$ 4.82 and its effective temperature to T$_{\text{pr}} \le 4000$ K. Our results suggest that SN~2024bch is a type IIn-L supernova that originated from a progenitor star consistent with a red supergiant. We show how the correct estimation of the mass-loss history of a supernova will play a major role in future multiwavelength observations.}


    \maketitle

\section{Introduction}

    Core-collapse supernovae (CCSNe) are the final product of the evolution of massive stars. They are among the most energetic phenomena in our Universe and are ideal sources for multiwavelength and multi-messenger studies, as notably proven by the detection of neutrinos from SN~1987A~\citep{Bionta1987, Hirata1987}. Despite the fact that many supernovae (SNe) have been extensively characterized from the radio to the X-ray band, with possible detections up to soft gamma rays \citep{Yuan2018, Xi2020, Chen2024}, the physical processes and mechanisms behind the collapse are not yet fully understood.

    Supernovae can be powerful cosmic ray (CR) accelerators and are believed to produce gamma rays up to the very high-energy (VHE) band (E $\ge$ 100 GeV; \citealt{Marcowith2018}). The interaction of the fast energetic SN ejecta with a dense and slower circumstellar medium (CSM) surrounding the progenitor triggers the formation of shock waves, which give rise to thermal optical, UV, and X-ray emission. Collisionless shocks are expected to accelerate a fraction of the charged particles up to high relativistic energies via the diffusive shock acceleration mechanism (e.g., \citealt{longair2011high, Murase2011, bell2013cosmic}), leading to nonthermal emission \citep{chevalier2016thermal}. In this scenario, proton--proton interactions between the accelerated protons and the CSM may give rise to VHE gamma rays \citep{Murase2024}. The VHE gamma-ray flux is expected to peak shortly after the explosion, when the SN ejecta encounter the innermost, densest region of the CSM, leading to an increase in proton--proton interactions \citep{Brose2022}.

    However, during the first tens to hundreds of days after the explosion, the possible VHE signal is significantly attenuated by the gamma-gamma absorption resulting from the interactions of VHE gamma rays with the soft optical photons emitted by the SN photosphere (e.g., \citealt{Tatischeff2009, Bykov2018, Cristofari2022, Brose2022}). However, a detailed characterization of this process is made difficult by the high number of degrees of freedom and degeneracy among the relevant parameters. Attenuation is influenced not only by the intrinsic properties of the SN itself, but also by the characteristics of the ejected material, the structure of the CSM (which is shaped by the progenitor star's mass-loss history), and the evolution of the photosphere, as its temperature and radius determine the population of low-energy target photons.

    To date, the dedicated VHE gamma-ray follow-up campaigns for different types of SNe performed by the current generation of Imaging Atmospheric Cherenkov Telescopes (IACTs) have not achieved any significant detections~\citep{Ahnen2017, Abdalla2019, Acharyya2023}. Nevertheless, the quest for gamma-ray signals from CCSNe remains an important goal for current and next-generation VHE facilities. In particular, future facilities such as the Cherenkov Telescope Array Observatory \citep[CTAO;][]{cta2018science}, with its foreseen improved sensitivity and wider accessible energy range, will likely be able to detect VHE counterpart of CCSNe up to $\sim$ 10 Mpc \citep[e.g.,][]{WhitePaper2021}, boosting our understanding of the physical mechanisms at work.

    The standard classification scheme of SNe groups different events based on their spectral features~\citep[e.g.,][and references therein]{Filippenko1997}. In particular, type II SNe exhibit hydrogen-rich spectra. These are the most common events and are typically associated with massive red supergiant (RSG) progenitors. Depending on the shape of their optical light curves, they were historically divided into type II-L (linear) and type II-P (plateau) based on the relative decay rate \citep{Barbon1979}. Although several studies have found evidence that the distribution of decay rates is a continuum (e.g., \citealt{Patat1994, Anderson2014, Gall2015, Sanders2015}), others suggest that a bimodal distribution can be seen when focusing on the decay rates of specific time ranges.  \citet{Li2011} and \citet{Faran2014} proposed a new method for discriminating between the two classes: the \textit{R}-band (\textit{V}-band) luminosity has to decay by 0.5 magnitudes or more between the peak of luminosity and 50 days after the explosion ($t_{50}$). 
       
    A further subtype of hydrogen-rich SNe is the type IIn. They exhibit spectra dominated by strong Balmer emission lines on a blue continuum, indicating a hydrogen-rich environment possibly caused by strong mass loss before the explosion~\citep{Schlegel1990}. These lines often include a narrow component -- with a full width at half maximum (FWHM) of about 100 km s$^{-1}$ -- from the un-shocked CSM and broader components (FWHM of several thousand km s$^{-1}$) arising from the interaction between the rapidly expanding ejecta and the slow, dense CSM \citep[e.g.,][]{Chevalier1994}. To form a CSM sufficiently dense to allow such narrow lines to persist for a long period after the explosion, the progenitor star must undergo significant mass loss prior to the explosion. Luminous blue variable (LBV) stars are often proposed as possible direct progenitors of type IIn SNe~\citep{Kotak2006, Trundle2008, Gal-Yam2009}. However, RSG candidates are also a widely considered possibility~\citep[e.g.,][]{Fransson2002, Smith2009, Mackey2014} despite likely not being major contributors, given the relative rates of type IIn SNe (e.g., \citealt{cold2023cosmic}) and the limited mass range from which RSGs originate.
   
    Due to the combination of high luminosity and strong CSM interactions, type IIn SNe are among the most likely CCSNe to exhibit a VHE emission component~\citep[e.g.,][]{Murase2024}. However, significant differences can exist between individual type IIn SNe due to different CSM properties, explosion parameters, and potential progenitor channels. For this reason, new subclasses have recently been suggested based on their photometric and spectroscopic evolution~\citep[]{Taddia2013, Habergham2014, Taddia2015}. Among them, IIn-L SNe are classified based on their similarity to SN~1998S \citep{chugai2001broad, fassia2001optical}. Their light curves exhibit a rapid type IIL-like decline, while their early-time spectral evolution resembles that of type IIn SNe. At later times, this evolves into a more typical IIL-like behavior, with broader lines. As in the case of SN~1998S, signatures of CSM interaction can persist at late times, indicating different episodes of mass-loss before the explosion \citep{pozzo2004source}. Overall, this behavior indicates a low-mass CSM~\citep[e.g.,][]{Smith2016}. In recent years, some SNe originally classified as type IIn have been reclassified as IIn-L under this newer definition~\citep[e.g.,][]{Faran2014, Kangas2016}. Their explosion sites have been found to be metal-rich, similar to those of normal type II-L and II-P SNe, supporting the scenario of massive RSG progenitors~\citep{Taddia2015}.
        
    SN~2024bch was discovered on January 29,  2024, at 06:02:50 UTC in the barred spiral galaxy NGC~3206 ($z$ = 0.003863; $\mu$ = 31.49 $\pm$ 0.45; $d = 19.9^{+3.8}_{-4.5}$ Mpc; \citealt{Tully2016, Castignani2022}), at coordinates RA(J2000) = 10h 21m 50.200s, Dec(J2000) = +$56^{\circ}$ $55^{\prime} \ 36.10^{\prime \prime}$~\citep{Wiggins2024}. It was initially classified as a type IIn due to its spectral similarity with SN~1998S at 3.5 days after the explosion~\citep{Balcon2024}. SN~2024bch has been studied extensively in the optical band and has been classified as a type II with CSM interaction~\citep{Tartaglia2024,Andrews2024}. We present VHE follow-up observations of SN~2024bch made with the CTAO Large-Sized Telescope prototype (LST-1) and optical follow-up taken with the Liverpool Telescope (LT). The latter allowed us to recover one spectrum at $\sim$~50 days after the explosion. We combined our VHE gamma-ray and optical observations to determine the general properties and the type of the event and its progenitor star. The spectrophotometric evolution in the optical band is crucial for constraining some of the parameters required to model the gamma-ray flux and reduce systematic uncertainties. This multiwavelength approach may play a significant role in correctly modeling observations with current and future IACTs. 
    
    The paper is structured as follows: We present our VHE gamma-ray and optical observations and analysis in Sects.~\ref{sec:2} and \ref{sec:3}, respectively. In Sect.~\ref{sec:4} we focus on constraining the physical parameters of the SN explosion, the SN ejecta, and the progenitor star. In Sect.~\ref{sec:5} we discuss how our results can be used to determine the class and progenitor star of SN~2024bch, while in Sect.~\ref{sec:new} we discuss how our data may be affected by gamma-gamma absorption. Finally, we outline our conclusions in Sect.~\ref{sec:6}.

    \section{VHE observations and data analysis}
    \label{sec:2}
   
    The LST-1 is the first 23~m diameter telescope that will form the core of the upcoming northern array of the CTAO \citep[CTAO-N;][]{cta2018science}. The telescope is located at the Observatorio del Roque de los Muchachos in the Canary Island of La Palma, Spain. Thanks to its large reflecting area of $\sim$ 400 m$^2$ and a camera constituted by high-quantum efficiency photomultiplier tubes, LST-1 is optimized for observations at the lowest energy edge of the VHE band, above a few tens of GeV~\citep{Abe2023b}.
    
    We triggered target of opportunity (ToO) observations on SN~2024bch with the LST-1 starting from February 13, 2024 (MJD 60353), about two weeks after its discovery. The follow-up began with a two-week delay because the gamma-gamma attenuation is expected to be strong during the first days after the explosion (see Sect.~\ref{sec:new}). The follow-up was extended for a total of six nonconsecutive nights collecting 14.6 hours of observations between February and March 2024 in the zenith range 28$^{\circ}$ -- 43$^{\circ}$. Observations were performed in the so-called wobble mode~\citep{Fomin1994} and in good atmospheric and in dark-to-low-moonlight conditions. The corresponding energy threshold is about $\sim 30$~GeV. 

    We followed a source-independent standard analysis scheme for SN~2024bch gamma-ray data, as described in~\citet{Abe2023}. After applying data quality cuts, based on pointing stability, level of constant night sky background and intensity of CRs, we selected 12.0 hours of good quality data. Data reduction is performed using the software package \texttt{cta-lstchain v0.10.7}, the standard analysis pipeline for LST-1 data~\citep{Lopez-Coto2021, lstchain-Zenodo_2024}. From the recorded shower images, random forests trained on Monte Carlo simulations, produced at a fixed declination of +61.66$^\circ$ (4.73$^\circ$ away from the SN~2024bch position), are used to determine the parameters of the incoming primary particles, energy, direction and \textit{gammaness}, a score for gamma-hadron separation. After determining the shower parameters, event selection cuts were applied and the data were processed to the final scientific level, where the response of the telescope was characterized using instrument response functions derived from Monte Carlo simulations assuming a point-like source hypothesis.

    For the 1D spectral analysis we used the open-source tool \texttt{Gammapy}~\citep{Donath2023, Acero2024}. We used one OFF region (i.e., a reflected region with respect to the ON region centered at the source position) to model and subtract the background contribution from our data, specifically the irreducible background from isotropic gamma rays and gamma-like hadronic air showers.

\begin{figure}
    \includegraphics[width=1\columnwidth]{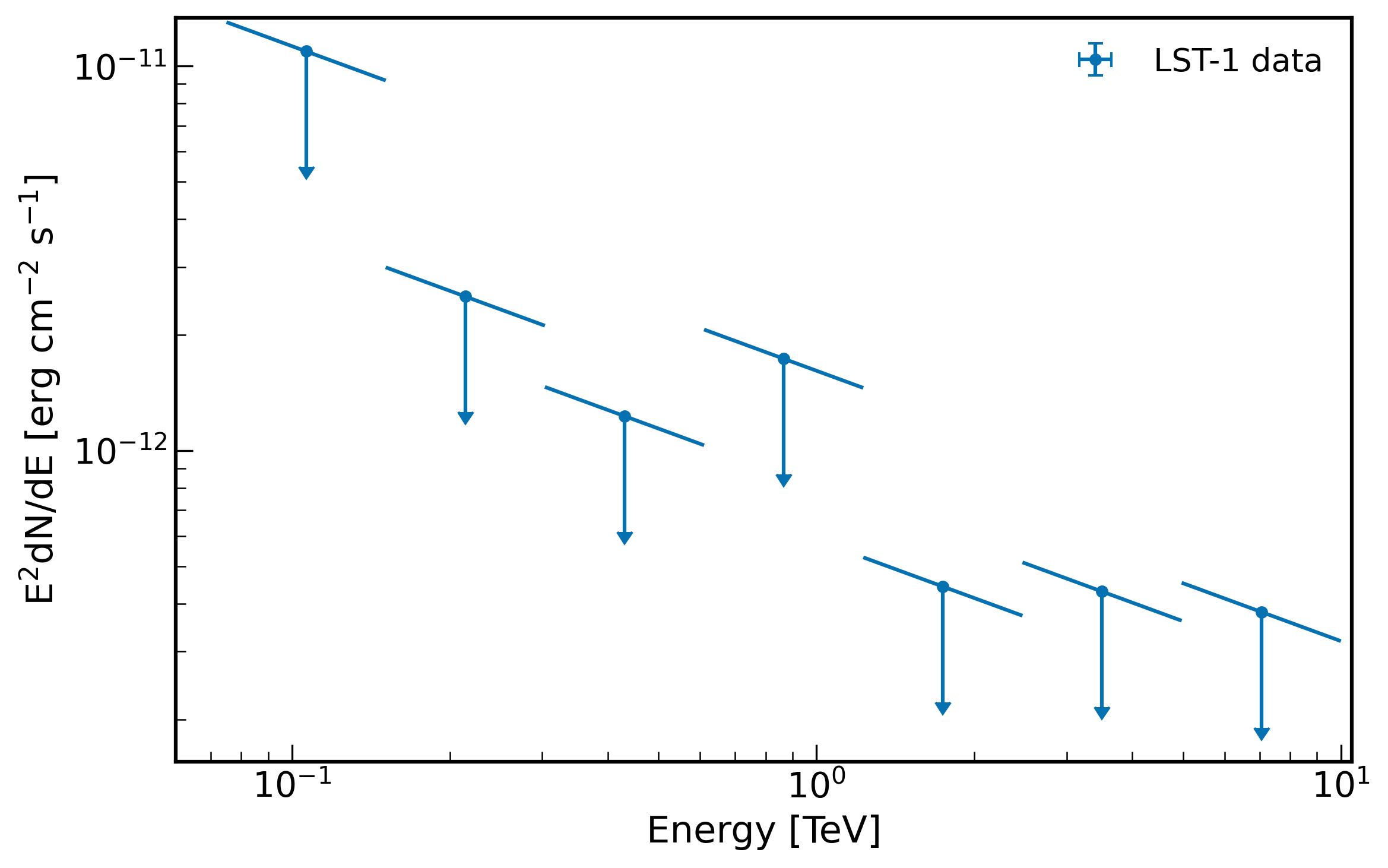}
    \caption{LST-1 differential ULs obtained from SN~2024bch observations between 75 GeV and 10 TeV, estimated with three energy bins per decade and assuming a fixed spectral index of -2.5 considering the full time-range. ULs are computed with a 2$\sigma$ confidence level.}
    \label{fig:sed_ul}
    \end{figure}

\begin{figure}[h!]
    \includegraphics[width=1\columnwidth]{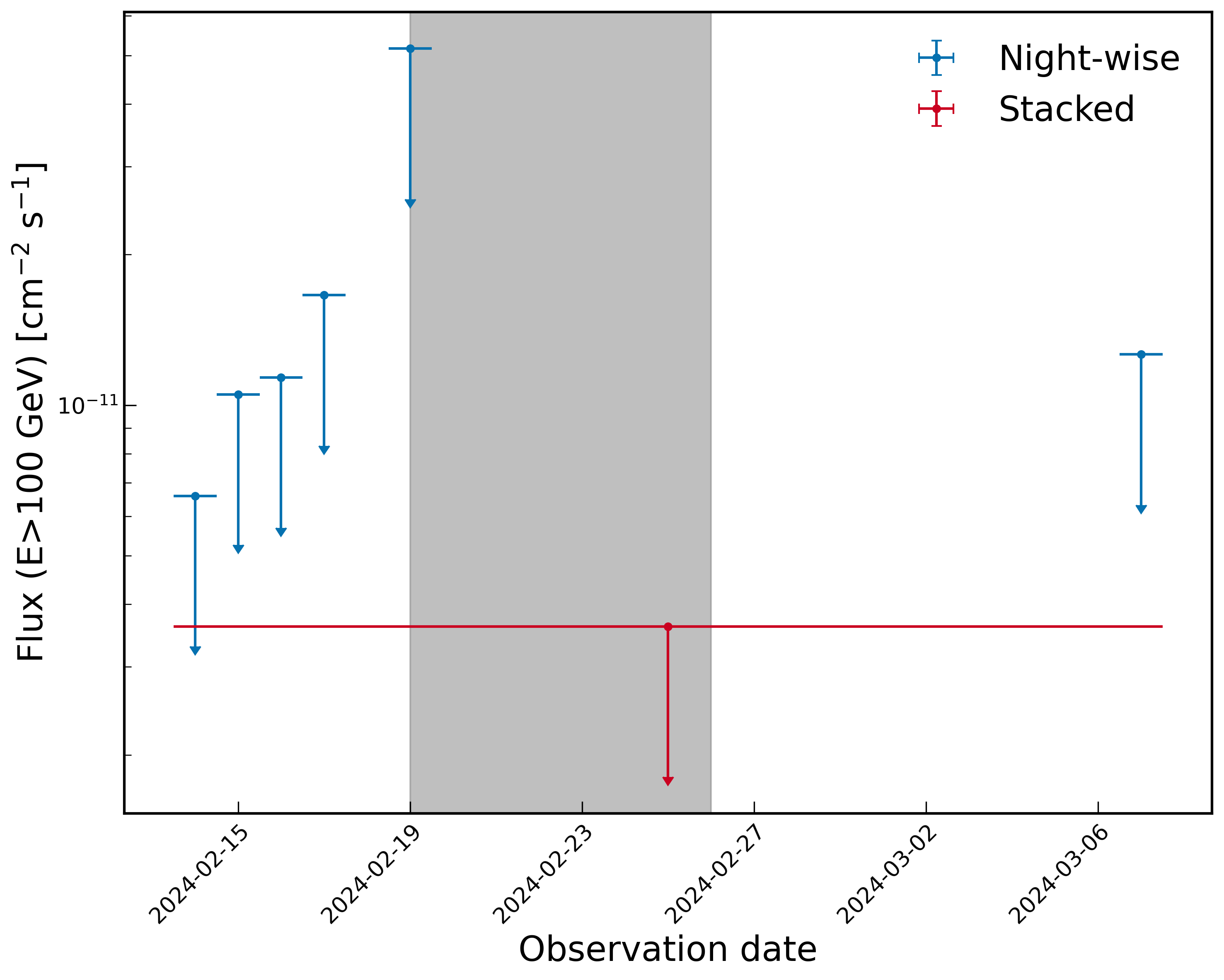}
    \caption{ULs for the gamma-ray light curves of SN~2024bch, computed between 100 GeV and 10 TeV. We compare the night-wise ULs (blue) with the integrated UL, obtained by stacking all results together (red). The gray-shaded area covers the nights with strong moonlight. ULs are computed with a 2$\sigma$ confidence level.}
    \label{fig:lc_ul}
    \end{figure}

    No significant gamma-ray excess was detected on any of the observed nights or from the full dataset. We computed upper limits (ULs) for the spectral energy distribution of SN~2024bch between 75 GeV and 10 TeV, assuming a single power law with a spectral index of $\Gamma = -2.5$ (see Sect.~\ref{sec:4.1}). The spectral energy distribution is shown in Fig.~\ref{fig:sed_ul}. Additionally, we calculated night-wise light curve ULs between 100 GeV and 10 TeV. Using the same spectral index, we also computed the integrated UL. We show our results in Fig.~\ref{fig:lc_ul} and Table~\ref{tab:ULs}.

\vspace{1cm}

    \section{Optical observations}
    \label{sec:3}

    \subsection{Light curves}
    \label{sec:3.1}
    
\begin{figure}
    \includegraphics[width=1\columnwidth]{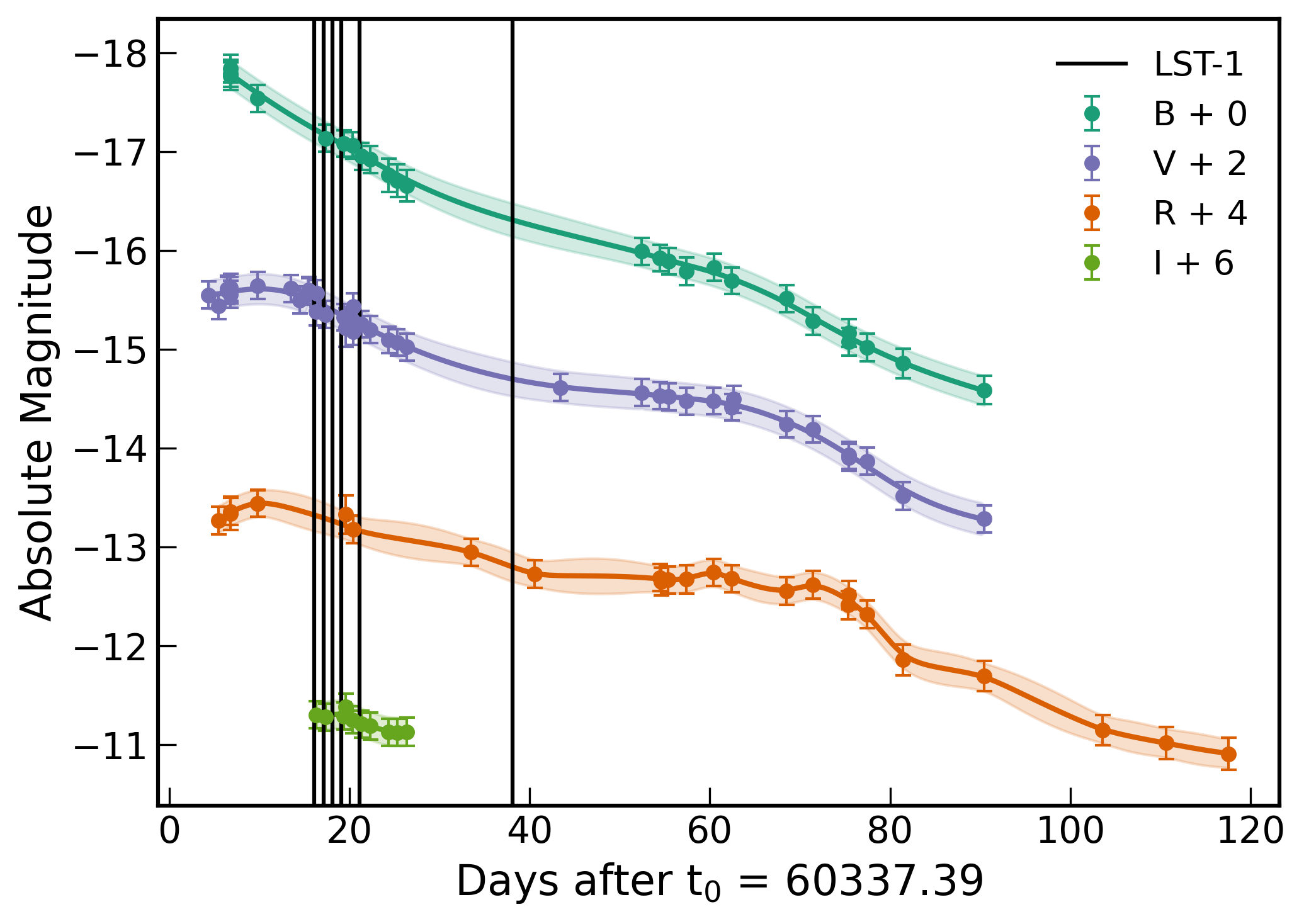}
    \caption{Public photometry of SN 2024bch (filled points). We used observations in the four filters \textit{B, V, R,} and \textit{I}. An offset is added to the data points for a better visualization. Solid lines are the result of GP interpolation obtained with \texttt{CASTOR} (see the main text). The uncertainties of the interpolated lines are shown as shaded contours. The vertical solid black lines show the LST-1 observation windows.}
    \label{fig:lc}
    \end{figure}

    We collected photometric data available in the American Association of Variable Star Observers (\texttt{AAVSO}) repository \citep{Kloppenborg2022} in four bands (\textit{B, V, R}, and \textit{I}). All magnitudes were converted into the relative standard photometric system and K-corrected. These light curves range between MJD 60341 and MJD 60455, covering the region around the peak and the plateau phase. The light-curve drop indicating the switch between plateau and nebular phase is only clearly visible in the \textit{R} band. We inferred a time of explosion of $t_0 = 60337.4 \pm 1.9$ MJD, which is adopted as reference time throughout the paper. Furthermore, the \textit{V} band peaks at $t_{\text{max}} = 60347.3 \pm 1.7$ MJD. We adopted a reddening along the line of sight of E$_{B-V}$ = 0.049, as estimated in \citet{Andrews2024} from Na ID line measurements. The distance of the event was assumed as the average distance of NGC~3206 from redshift-independent estimates from NASA Extragalactic Database (NED)\footnote{\url{https://ned.ipac.caltech.edu/}}. We find $d = 17.6 \pm 1.1$ Mpc, corresponding to a distance modulus of $\mu = 31.23 \pm 0.13$. All photometric data used in this analysis are shown in Fig.~\ref{fig:lc}, compared with the LST-1 observing nights.

\begin{figure*}
    \includegraphics[width=1\textwidth]{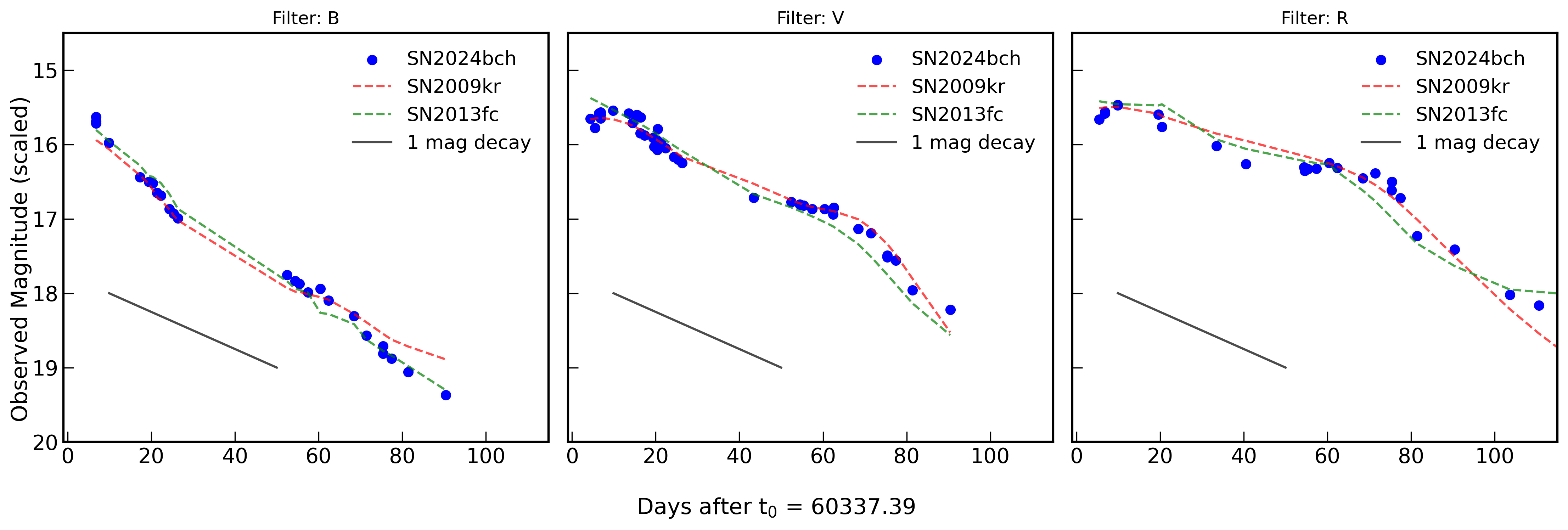}
    \caption{Comparison of the \textit{BVR} light curves of SN~2024bch (blue dots) with those of SN~2009kr and SN~2013fc (dashed red and green curves, respectively). The data for SN~2009kr and SN~2013fc are scaled to match SN~2024bch and interpolated over the same time interval using GP interpolation (see the main text for details). Additionally, we show a reference curve representing a decay of 1 magnitude between $t_{\text{max}}$ and $t_{50}$. }
    \label{fig:comparison}
    \end{figure*}

    We used the open-access software \texttt{CASTOR v2.0} \citep{Simongini2024, castor_zenodo_2025} to interpolate our light curves via Gaussian process (GP) regression techniques and compared them to a catalog of 150 CCSNe via chi-square test. We found this SN to be most similar to SN~2009kr \citep{Elias-Rosa2010}, SN~2013fc \citep{Kangas2016}, and SN~2008fq \citep{Faran2014}, as shown in Fig.~\ref{fig:comparison}. These CCSNe were all classified as type II, with linear decaying light curves and 1998S-like spectra. The interpolated \textit{V}-band light curve of SN~2024bch decays by 1.05 $\pm$ 0.11 magnitudes between the peak of luminosity and $t_{50}$, while the \textit{R}-band light curve decays by 0.74 $\pm$ 0.10 magnitudes (see Fig.~\ref{fig:comparison}). According to the criteria determined by \citet{Li2011} and \citet{Faran2014}, and to the statistical comparison performed with \texttt{CASTOR}, SN~2024bch shows a clear linear decay tens of days after the explosion, similar to a common type II-L.  

    \subsection{Spectra}

\begin{figure*}
    \includegraphics[width=1\textwidth]{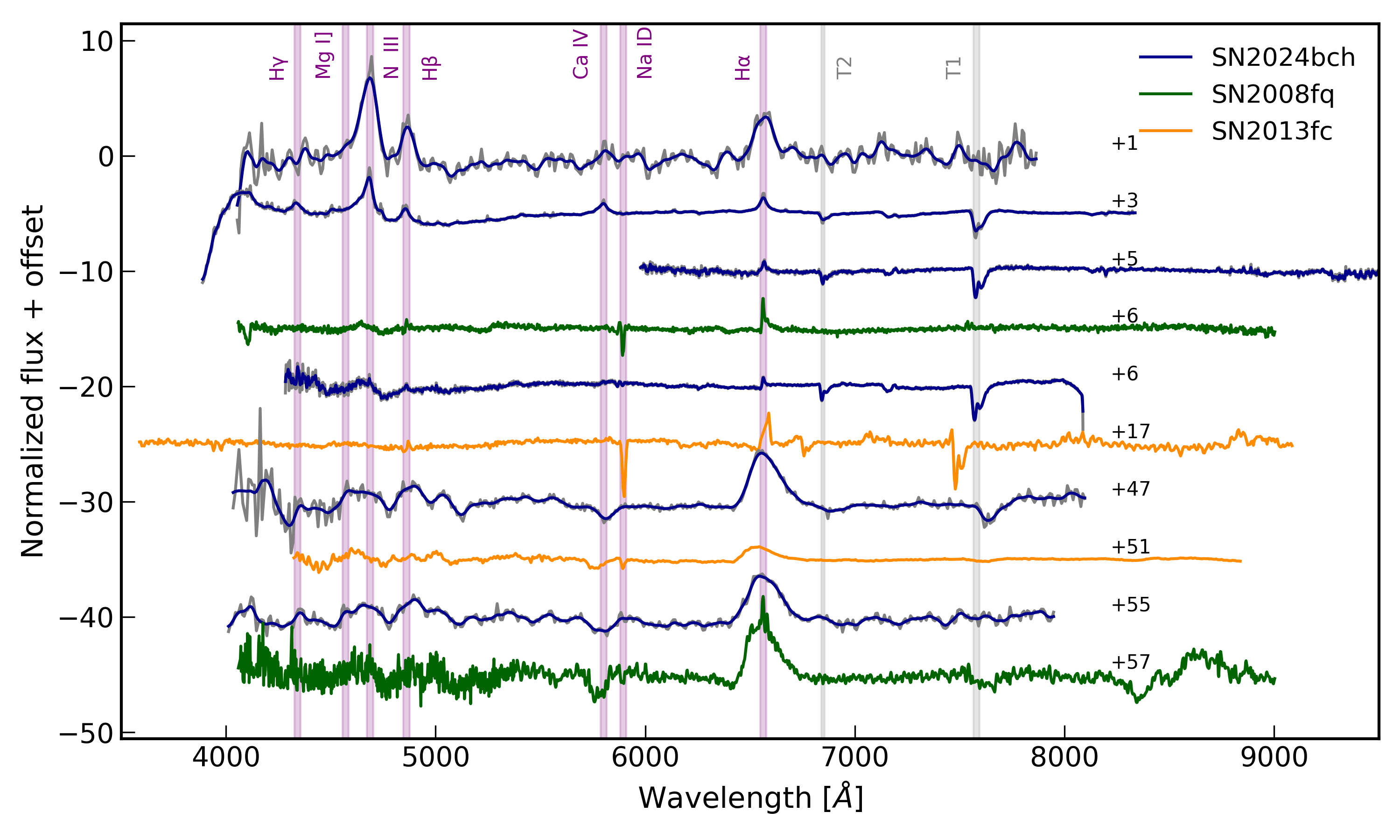}
    \caption{Spectra of SN~2024bch compared with two spectra of SN~2013fc (at 17 and 51 days post-explosion) and two spectra of SN~2008fq (at 6 and 57 days post-explosion). We removed the continuum from every spectrum and normalized to order zero. We added an offset for better visualization. The main spectral features of SN~2024bch are highlighted as purple-shaded vertical bands. Telluric lines are identified as "T1" and "T2" (gray bands). The spectrum taken with LT is the second from the bottom of the SN~2024bch spectra, at 47 days post-explosion. The other spectral data for the three SNe are taken from \texttt{WISeREP} \protect\citep{Yaron2012}.}
    \label{fig:sp}
    \end{figure*}

    SN~2024bch was observed on a single night by the LT~\citep{steele2004liverpool} via a ToO observation. The LT is a 2-m robotic Cassegrain reflector, located at the same site as the LST-1. Low-resolution ($\textit{R} = 350$) spectroscopy was performed on MJD 60385.018, 47.6 days after the explosion with the SPRAT spectrograph~\citep{Piascik2014}. We obtained one spectrum in the range 4000--8000~\AA \, as shown in Fig.~\ref{fig:sp}. Data were reduced using the standard LT pipeline. Our LT spectrum is complemented with ancillary low-resolution spectra collected from \texttt{WISeREP}~\citep{Yaron2012}. All spectra were taken between 4000 and 9000 \AA \, in the host-galaxy reference frame and thus are not redshift-corrected. We cut every spectrum between 4000 and 8300 \AA \, to avoid data with low signal-to-noise ratios (S/Ns), effectively removing the near-IR region of the spectrum. Figure~\ref{fig:sp} shows the temporal evolution of SN 2024bch spectra between 1 and 55 days after the explosion. The main spectral features are highlighted with vertical lines. At early times (8, 6, 4, and 3 days before the peak of maximum luminosity, MJD 60347.19, $t_0 + 9$ days), the spectra consist of a blue, almost featureless continuum. The only discernible lines, apart from the telluric lines around 7580 \AA \, and 6848 \AA \,, are the Balmer lines at 6563 \AA \,, 4861 \AA, and 4341 \AA \,, the C IV line at 5801 \AA, \, and the N III line at 4687 \AA. By mid-late times (35 and 45 days after the peak), the blue continuum has cooled. The H$\alpha$ is considerably broader and stronger and has remained the most distinguishable feature along with the H$\beta$ line, which is weaker and broader. The N III line at 4687 \AA\, is less discernible and possibly blended with Mg I] line at 4571 \AA. The C IV line at 5801 \AA \, has been substituted by a Na ID line at 5895 \AA \,. \\ 
    
    The characteristic H$\alpha$ profile appears to undergo a significant change in morphology between early and late phases. We decomposed the line into a Lorentzian and a Gaussian profile at early times (respectively for the narrow and the broad components), and into two Lorentzian at late times. The first available spectrum (at 1 day after the explosion) presents two-peaks around the Balmer line, possibly due to contamination of other nature, as this behavior is not shown in the following days. In this case, we used only a Gaussian distribution to fit the single peak around the line. For each line we estimated the amplitude with respect to the continuum and normalized it by a factor of $10^{15}$ erg s$^{-1}$ cm$^{-2}$ \AA$^{-1}$, the FWHM, the relative velocity and the luminosity. At early times, the H$\alpha$ profile becomes weaker and slower as it gets closer to the maximum of luminosity and there are signs of P-Cygni profiles. From 1 to 6 days after the explosion ($-$8 to $-$3 days before the peak), the amplitude changes from $\sim$ 4 to $\sim$ 1, the FWHM changes from $\sim 70$ \AA \, to $\sim$ 10 \AA, the velocity from $\sim$ 3000 km s$^{-1}$ to $\sim$ 500 km s$^{-1}$ and the luminosity from $\sim$ 9 $\times 10^{39}$ erg s$^{-1}$ to $\sim$ 4 $\times 10^{38}$ erg s$^{-1}$. On the contrary, at late times the H$\alpha$ profile has become significantly stronger and broader, with amplitude on the order of 5, FWHM on the order of 150 \AA, velocity on the order of 7000 km s$^{-1}$ and luminosity on the order of $2\times 10^{40}$ erg s$^{-1}$. Furthermore, no signs of P-Cygni profiles are found in our last two spectra. 

    We again used \texttt{CASTOR} to interpolate our spectra with 2D GPs. In particular, we built 60 synthetic spectra from 0 to 60 days after the explosion, ranging from 4000-8390\AA. Note that differently from what is described in \citet{Simongini2024}, we did not use photometric points from light curves to build synthetic spectra, but we used only the available spectra of SN~2024bch shown in Fig.~\ref{fig:sp}, effectively interpolating observed spectra in time and wavelength. We determined the redshift from the mean relative shift of the absorption lines in the synthetic spectra, obtaining $z = 0.00386 \pm 0.00016$. This value agrees with the latest redshift estimation for NGC~3206~\citep{Castignani2022} and reported also in~\citet{Balcon2024}. Similarly, we estimated the expansion velocity at $t_0$ and at $t_{\text{max}}$ from the average half-width of the spectral lines. We obtain $V_{\text{sh}} = 3088 \pm 50 $ km s$^{-1}$ and $V_{p} = 4688 \pm 50$ km s$^{-1}$, respectively. In the following analysis, we assume that $V_{\text{sh}}$ and $V_{\text{p}}$ represent the shock velocity and the ejecta velocity, respectively.
    
    \section{Physical parameters}\label{sec:4}
    
    \subsection{Gamma-ray constraints}\label{sec:4.1}

    Following the work from \citet{Abdalla2019}, we used the semi-analytical model for CCSNe described in \citet{Dwarkadas2013} in order to place our photon flux ULs into a physical context. This model assumes a constant stellar mass-loss rate $\dot M$ and wind velocity $u_\text{w}$, which is usually known as the "steady-wind" scenario described by \citet{Chevalier1982}. Under this assumption, the CSM density is given by the mass continuity equation: 
    \begin{equation}
        \rho_{\text{CSM}} = \frac{\dot M}{4\pi u_\text{w} r^{2}},
    \end{equation}
    where $r$ is the outer radius of the CSM. For CCSNe, the model of \citet{Dwarkadas2013} gives the following relation of the expected gamma-ray flux as a function of stellar mass-loss parameters, SN explosion properties, and time passed since the explosion:
    \begin{equation} 
    F_\gamma(t) = \frac{3 q_\alpha \xi \kappa C_1 m^3}{32\pi^2 (3m-2) \beta\mu m_\text{p}} \left(\frac{\dot M}{u_\text{w}}\right)^2 \frac{1}{d^2} t^{m-2}, 
    \end{equation} 
    where $q_\alpha$ is the gamma-ray emissivity normalized to hadronic CR energy density, $\xi$ is the fraction of the shock energy flux converted into CR proton energy, $m$ is the expansion parameter, $\beta$ is the fraction of total volume already shocked, $\mu$ is the mean molecular weight of the nuclear targets in the CSM, $d$ is the distance of the event, $\kappa$ is the ratio of the forward shock radius to the contact discontinuity, $C_1$ is a constant that can be expressed in terms of the geometry of the explosion, and $m_\text{p}$ is the proton mass. Following the same procedure as \citet{Abdalla2019}, we substituted $\kappa C_1$ with $V_{\text{sh}}/(m \ t^{m-1}),$ where $V_{\text{sh}}$ is the shock velocity, which led to the following relation: 
    \begin{equation} 
    F_\gamma(t) = \frac{3q_\alpha \xi V_{\text{sh}} m^2}{32 \pi^2 (3m-2) \beta\mu m_p} \left(\frac{\dot M}{u_\text{w}}\right)^2 \frac{1}{d^2} \frac{1}{t}.
    \label{eq:due}
    \end{equation} 
    Moreover, following their prescriptions, we set $\xi$ = 0.1, $\mu$ = 1.4, $\beta$ = 0.5 and $m$ = 0.85. Assuming a spectral index of $\Gamma = -2.5$, the gamma-ray emissivity normalized to the CR energy density above 100 GeV is q$_\alpha (>100 \,\mbox{GeV}) = 9.5 \times 10^{-18}$ cm$^{3}$ erg $^{-1}$ s$^{-1}$ \citep{Drury1994}. Assuming the shock velocity   (assuming free expansion) and all the explosion and CSM parameters are constant, we can invert Eq.~\ref{eq:due} to convert our gamma-ray flux ULs to mass-loss-rate to wind-velocity ratio ULs: 

    \begin{equation} 
    \left(\frac{\dot M}{u_\text{w}}\right) \propto \left(\frac{F_\gamma t}{V_{\text{sh}}}\right)^{1/2} d,
    \end{equation}where the distance ($d$) and the shock velocity ($V_{\text{sh}}$) are constrained from optical data. It is important to note that a change in the assumed spectral index has an impact on the gamma-ray emissivity. By using a spectral index of $\Gamma$ = $-2$, we obtain a q$_\alpha$($>$ 100 GeV) = $1.3\times 10^{-16}$ cm$^{3}$ erg $^{-1}$ s$^{-1}$, accounting for about a factor of 10 of difference and implying an effect on the estimated mass-loss rate wind velocity ratio limited to a factor of $\sim$ 3.
    Table~\ref{tab:ULs} shows the night-wise ULs evaluated between 100 GeV and 10 TeV and the corresponding luminosity ULs derived with the adopted value of distance. The mass-loss rate and the wind velocity are pre-explosion parameters and they remain constant after the explosion. Therefore, we computed the reference mass-loss wind velocity ratio UL by using the stacked gamma-flux UL and using the exposure-weighted average time ($t_{\text{ref}} = 19.89$ days) as the reference time. Note that the assumed model does not take into account the gamma-gamma absorption, which represents the main source of attenuation of VHE photons. See Sect.~\ref{sec:new} for a detailed discussion about this approximation. 
    
    \begin{table*}
    \centering
    \caption{Results of the LST-1 observations.}
    \label{tab:ULs}
    \begin{tabular}{ll|cccc} 
    \toprule 
    \toprule 
    Night  & Phase\tablefootmark{a} & Flux UL & Luminosity UL & $\dot M/u_\text{w}$ UL  & Exposure\\ 

    [MJD]   & [days] & [$10^{-11}$ cm$^{-2}$ s $^{-1}$] & [$10^{42}$ erg s $^{-1}$] & [$10^{-4}\frac{\text{M}_\odot}{\text{yr}}\frac{\text{s}}{\text{km}}$] & [h] \\ 
    
    \midrule 
    60353 & 16 & 0.66 & 0.24   &  1.47 & 3.52\\ 
    60354 & 17 & 1.05 & 0.39   &  1.92 & 2.72\\ 
    60355 & 18 & 1.14 & 0.42   &  2.05 & 2.83\\ 
    60356 & 19 & 1.66 & 0.61   &  2.55 & 1.57\\ 
    60358 & 21 & 5.16 & 1.91   &  4.72 & 0.48\\ 
    60375 & 38 & 1.26 & 0.47   &  3.14 & 1.46\\ 
    \midrule   
          &         & 0.361 & 0.133  &   1.21 & 12.58\\ 
    \bottomrule
    \end{tabular}
    \tablefoot{Photon flux above 100 GeV, luminosity and mass-loss rate wind velocity ratio ULs estimated for each of the six nights of observation with the LST-1. The last row presents the overall results: the photon flux stacked over the six nights of observations and the corresponding luminosity, the UL on the mass-loss rate and wind velocity ratio, obtained by averaging the results from the six nights while weighting by the exposure time and reference time. \tablefoottext{a}{Phases are expressed with respect to the estimated time of explosion $t_0$ = 60337 MJD. }
    }
    \end{table*}

    \subsection{Optical constraints}
    \label{sec:par}

\begin{figure}
    \includegraphics[width=1\columnwidth]{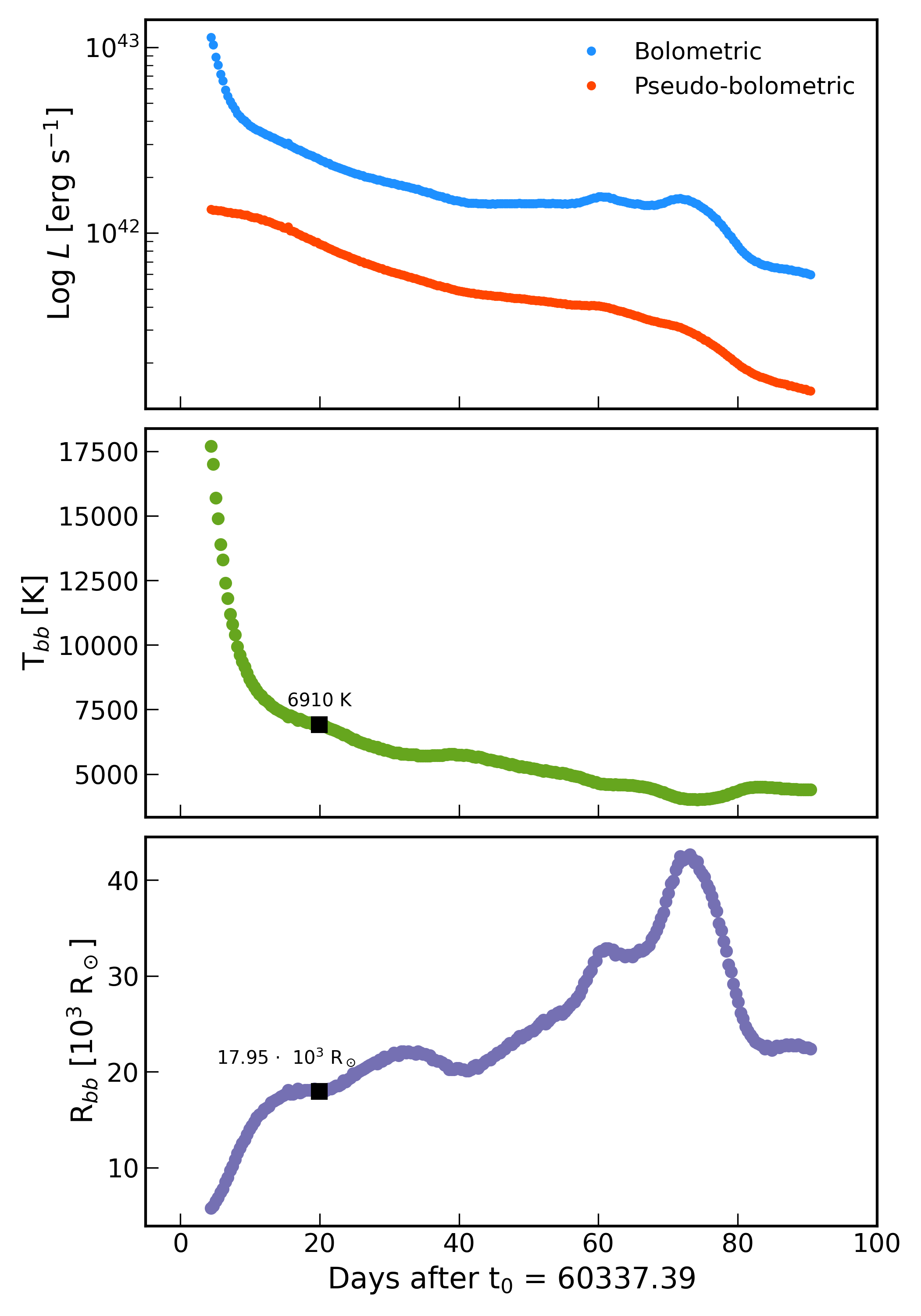}
    \caption{Evolution of the bolometric and pseudo-bolometric luminosity (top panel), photospheric temperature (central), and photospheric radius (bottom) of SN~2024bch derived with \texttt{SuperBol}. We highlight the values of temperature and radius reached at $\sim$ 20 days after the explosion. 
    }
    \label{fig:phot}
    \end{figure}

    The optical spectrophotometric evolution of a SN allows one to characterize several properties of the event and constrain the nature of its progenitor star. Moreover, optical constrains reduce the uncertainty in modeling the gamma-ray flux and provide tools for understanding the physical processes at play. We used the open-access software \texttt{SuperBol} \citep{Nicholl2018} to derive the evolution of the photospheric temperature ($T_{\text{bb}}$) and radius ($R_{\text{bb}}$) of SN~2024bch and the bolometric and pseudo-bolometric luminosity based on our interpolated BVR light curves (see Sect.~\ref{sec:3.1}). Photospheric parameters were estimated by assuming a black-body law of emission of the photosphere. We show our results in Fig.~\ref{fig:phot} and Table~\ref{tab:results}. The bolometric luminosity peaks at $L_{\text{bol}} = (1.1 \pm 0.7)\times 10^{43}$ erg s$^{-1}$ on day 4.4, while the pseudo-bolometric luminosity (which accounts only for the \textit{BVR} contribution) peaks at $L_\text{p} = (1.3 \pm 0.1) \times 10^{42}$ erg s$^{-1}$. Note that the time of peak luminosity is coincident with the first available epoch, suggesting that the real peak could have happened in previous epochs, as shown in \citet{Andrews2024}. Similarly, the evolution of the photospheric temperature exhibits a peak at 17700 K, followed by a steep decline in the next days, possibly related to the disappearing of narrow features. During the bulk of the LST-1 observing nights the temperature has fallen to a value of about $\sim$ 7000 K, reaching almost $6000$ K on day 40. Finally, the photospheric radius reaches a value of $\sim 18 \times 10^3 R_\odot$ at the time of the LST-1 bulk of observations and keeps expanding during the entire plateau phase reaching a maximum of $42.65 \times 10^3 R_\odot$ on day 73. Coincidentally with the light curve drop, after the peak the radius has a steep decay. 

    All the other parameters shown in Table~\ref{tab:results} are obtained with \texttt{CASTOR v2.0}, based on the luminosity curve derived with \texttt{SuperBol}. Parameters are estimated applying analytical models directly on the luminosity curve, assuming the most general physical framework, such as spherical and isotropic explosion, perfect adiabacity at peak luminosity and a modified black-body law of emission. The energy partition is based on the SN~1987A model, with a canonical 99.9$\%$ of the total energy transported by neutrinos and 0.1$\%$ by photons. For a detailed overview on how every parameter is derived and the inter-degeneracies between parameters, see \citet{Simongini2024}.
    
    Among the estimated parameters, this analysis allows us to constrain the mass and the radius of the progenitor star. The progenitor's mass is derived from the mass of the ejecta, assuming perfect conservation of mass at explosion and assuming two possible values for the remnant, effectively creating an interval of possible values. We obtained $M_{\text{pr}} = 11$ -- $20\,M_\odot$. This interval overlaps with previous estimations from \citet{Tartaglia2024}, who reported a possible progenitor mass in the range $15$ -- $18 \,M_\odot$, and \citet{Andrews2024} who assumed a value of $15 \,M_\odot$ to model the spectral evolution of the SN. Taken together with previous studies, our result supports the conclusion that the progenitor mass was likely around $15 M_\odot$. On the other hand, the progenitor's radius is constrained using the scaling relation from \citet{Shussman2016}: 
    \begin{equation}
        \text{ET} \approx 0.1 M_{\text{ej}} V_{\text{p}} R_{\text{pr}}
    ,\end{equation}
    where ET is time-weighted integrated bolometric luminosity (e.g., \citealt{Katz2013, Nakar2016, Zimmerman2024}) and $M_{\text{ej}}$ is the mass of the ejecta. We obtain $R_{\text{pr}} = 531 \pm 125 R_\odot$, similar to the assumed value of $R_{\text{pr}} = 501 R_\odot$ from \citet{Andrews2024}. 
        
    \begin{table}
    \centering
    \caption{Parametric map of SN~2024bch: general parameters of the event, properties of the ejecta, properties of the photosphere, and main parameters of the progenitor star.}
    \label{tab:results}
    \begin{tabular}{lc} 
     \toprule
     \toprule
     Parameter        &    Value      \\
     \midrule 

     $t_0$ (MJD)            &     60337.4 $\pm$ 1.9       \\ 
     $t_{\text{max}}$ (MJD)      &     60347.3 $\pm$ 1.7        \\
     $z$                           &     0.00386 $\pm$ 0.00016     \\ 
     $d$\tablefootmark{a} (Mpc)    &     17.58 $\pm$ 1.09          \\ 
     $E_{\text{B-V}}$\tablefootmark{b}         &     0.049                     \\ 

     \midrule 

     $V_{\text{sh}}$ (km s$^{-1}$)       &    3088  $\pm$ 50              \\ 
     $V_{\text{p}}$\tablefootmark{c} (km s$^{-1}$)   &    4688 $\pm$ 50               \\ 
     $M_{\text{ej}}$ ($M_{\odot}$)   &    9.72 $\pm$ 2.24             \\ 
     $E_{\text{tot}}$ (10$^{51}$ erg)   &    1.27 $\pm$ 0.27             \\ 
     $M_{\text{Ni}}$ ($M_{\odot}$)       &    0.032 $\pm$ 0.008           \\             

    \midrule 

    $L_{\text{pseudo}}$\tablefootmark{c} (10$^{41}$ erg s$^{-1}$ ) & 13.4 $\pm$ 1.0 \\ 
    $L_{\text{bol}}$\tablefootmark{c}        (10$^{43}$ erg s$^{-1}$ ) &  1.1 $\pm$ 0.7 \\ 
    $T_{\text{bb}}$\tablefootmark{d}     (10$^3$ K)                &  17.70 $\pm$ 3.02 \\ 
    $R_{\text{bb}}$\tablefootmark{d}          ($10^3$R$_{\odot}$)        &  43.31 $\pm$ 7.87 \\

    \midrule 

    $M_{\text{pr}}$ ($M_{\odot}$)      & $11-20$ \\ 
    $R_{\text{pr}}$ ($R_{\odot}$)    & 531 $\pm$ 125 \\ 
    $L_{\text{pr}}$ ($\mbox{log}L/L_\odot$)    & $3.81-4.82$ \\ 
    $T_{\text{pr}}$ (K)            & $2250-4023$ \\ 

    \bottomrule
    \end{tabular}
    \tablefoot{
    \tablefoottext{a}{Average redshift-independent estimate from NED (see the main text);}
    \tablefoottext{b}{From \protect\citet{Andrews2024}}; 
    \tablefoottext{c}{Estimated at the time of maximum luminosity; }
    \tablefoottext{d}{Maximum value. }
    }
    \end{table}

    \subsection{Pre-explosion image}

\begin{figure}
    \includegraphics[width=1\columnwidth]{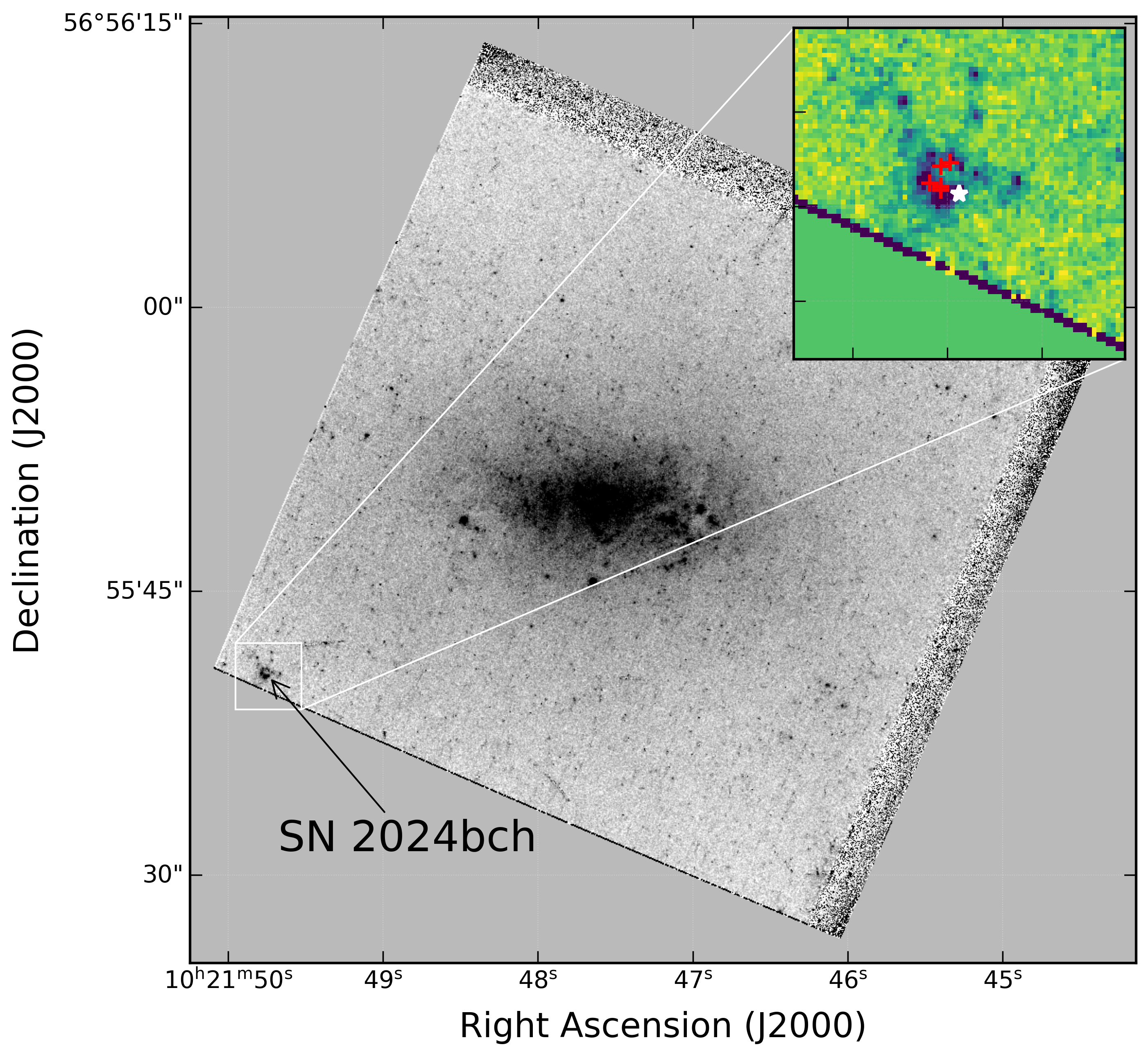}
    \caption{Location of SN~2024bch in its host galaxy, NGC~3206. This pre-explosion image was taken by the \textit{Hubble} Space Telescope on May 14, 2001, almost 20 years before the explosion. We zoom in on the region around the coordinates of SN~2024bch (white star), where different stars are identified as candidate progenitors (red crosses), and use a different color-map to highlight the S/N. 
    }
    \label{fig:su}
    \end{figure}

    We analyzed one image of the host-galaxy of SN~2024bch prior to the explosion, taken with the \textit{Hubble} Space Telescope\footnote{\url{https://hla.stsci.edu/hlaview.html}} on May 14, 2001 (Fig.~\ref{fig:su}). The region around the coordinates of the explosion appears more luminous than the background, significantly above the magnitude limit of the map. Table \ref{tab:luminosity} collects the properties of the candidate progenitor stars identified by different photometric catalogs. The small differences in the coordinates might be attributed to the different techniques used and astrometric errors. Therefore, we assumed that all the catalogs identify the same candidate. However, even if they correspond to different sources, our estimations remain valid. The apparent magnitudes are converted into luminosity using the host-galaxy distance and rescaled with respect to the solar luminosity with a base-10 logarithm. Hence, the different catalogs identify a candidate star with a luminosity at most between $L_{\text{pr}} = 10^{3.81}$ and $10^{4.82}$ $L_\odot$. Using the Stefan-Boltzmann law, this luminosity interval corresponds to an effective temperature in the range between $T_{\text{pr}} = 2000$ and $4000$ K.

    \begin{table}[h!]
    \centering
    \caption{Pre-explosion parameters of the candidate progenitor of SN~2024bch.
    }
    \label{tab:luminosity}
    \begin{tabular}{l|cccc} 
    \toprule
    \toprule
    Catalog & RA [deg] & Dec [deg] & MAG & Log($L/L_\odot$) \\
    \midrule 
    DAOphot             &       155.45741 &     56.92795  & 21.59  & 3.91 \\ 
    DAOphot             &       155.45735 &     56.92793  & 21.57  & 3.92 \\ 
    SExtractor  &       155.45735 &     56.92800  & 20.50  & 4.35 \\ 
    GSCII               &       155.45735 &     56.92793  & 19.76  & 4.65 \\ 
    HSC             &   155.45730 &     56.92801  & 21.84  & 3.81 \\ 
    PS1                 &       155.45735 &     56.92794  & 19.32  & 4.82 \\ 
    \bottomrule
    \end{tabular}
    \tablefoot{For each catalog, we report the coordinates, the apparent magnitude and the base-10 logarithm of the ratio of the luminosity to the solar luminosity ($L_\odot = 3.826 \times 10^{33}$ erg s$^{-1}$).}
    \end{table}

\section{On the supernova type and the progenitor star} 
\label{sec:5}

    After the luminosity peak, SN~2024bch light curves decay linearly within the following $\sim$ tens of days, in good agreement with the definition of type II-L SNe. We found this behavior to be similar to the ones showed by SN~2009kr, SN~2013fc, and SN~2008fq. Apart from SN~2009kr, which was classified as a II-L~\citep{Elias-Rosa2010}, SN~2013fc and SN~2008fq belong to the subgroup of type IIn-L~SNe \citep{Taddia2013, Kangas2016}. 

    We took the adopted values of reddening and distance into account to determine the absolute magnitudes (see Fig.~\ref{fig:lc}). The absorption relative to every filter is obtained using the prescriptions from \citet{McCall2004}. The \textit{V} band rises to its peak value, $-$17.64 $\pm$ 0.13, within the first 10 days after the explosion. In the same period, the \textit{B} band reaches a value of $-$17.84 $\pm$ 0.14 and the \textit{R} band a value of $-17.44 \pm 0.13$. The quick rise is similar to that of other IIn objects (e.g., SN~1998S, SN~2007pk, and SN~2008fq), as is the \textit{V}-band peak brightness~\citep[$-$18.5 $<$ V$_{\text{max}}$ $<$ $-$19.3;][]{Taddia2013} despite being under-luminous. Nevertheless, the \textit{B}-band maximum aligns with the average value of type II-L \citep[$-$17.98 $\pm$ 0.34;][]{Richardson2014}. Note that our values are slightly lower than the ones reported in \citet{Andrews2024} due to the different adopted distance. Moreover, as shown in Fig.~\ref{fig:lc}, the plateau lasts for about $\sim$ 70 days, consistent with the low end of the II-L distribution reported in \citet{Valenti2015}. The early-time spectra exhibit weak signatures of CSM interaction in the form of narrow and weak H$\alpha$ profiles, which seem to disappear soon after the peak of luminosity in favor of broad and strong lines at later times. Furthermore, SN~2024bch was originally classified as IIn due to the spectral similarity to SN~1998S using the classification software \texttt{GELATO}~\citep{Harutyunyan2008}. Hence, the overall spectro-photometric behavior of SN~2024bch resembles that of CSI interacting type IIn-L SNe. 
    
    We found for the progenitor star $\text{M}_{\text{pr}} = (11-20)\, M_\odot$, $\text{R}_{\text{pr}} = 531 \pm 125 \, R_\odot$, $\text{L}_{\text{pr}} \le 10^{4.82}\, L_{\odot}$, and $\text{T}_{\text{pr}} \le 4000\, \text{K}$ from optical modeling and pre-explosion images and $\dot M \le 10^{-3}$ $M_\odot$ yr$^{-1}$ assuming $u_\text{w}$ = 10 km s$^{-1}$ from gamma-ray studies. According to~\citet{Taddia2015}, who studied the metallicity levels in the explosion sites of different CSI SNe, the progenitor stars of type IIn-L SNe are more likely RSGs, as for II-P and II-L SNe. These are among the most common progenitors for type II SNe. Typical RSGs have masses between 8 and 20 $M_\odot$, radii up to 1500 $R_\odot$ and luminosities between $10^4$ and $10^5$ $L_\odot$ for the brightest objects~\citep{Smartt2015}. The different rate of occurrence between II-L and II-P events suggests that the progenitors of II-L are somehow more massive and thus more luminous than those of II-P progenitors~\citep{Branch2017}. Moreover, RSGs have relatively low wind velocity (u$_\text{w}$ $\sim$ 10 km s$^{-1}$) and mass loss rates in the range $10^{-6}$ -- $10^{-4}$ $M_\odot$ yr$^{-1}$. \citet{Mackey2014} proposed the possibility of RSGs being progenitors for CSI SNe in the presence of nearby hot stars. The companion star can photo-ionize up to 35$\%$ of the gas lost by the RSG during its lifetime, forming a dense and confined shell around it. When the star explodes, the ejecta will interact with the dense shell, producing the characteristic features of type IIn SNe. Based on the result achieved by our analysis, we might infer that the progenitor star parameters agree with the RSG scenario, as also supported by the spectro-photometric behavior. This scenario was also explored by \citet{Andrews2024} who, assuming a progenitor star of M = 15 $M_\odot$, found a wind velocity of 35 -- 40 km s$^{-1}$ and a mass-loss-rate of $\dot M$ = $10^{-3} - 10^{-2}$, consistent with our estimate.
    
    However, the LBV scenario cannot be ruled out, as LBVs are most likely to be progenitors for IIn explosion~\citep[e.g.,][]{Gal-Yam2009, Mauerhan2013}. LBVs are among the most luminous stars known ($L \sim 10^{5}$--$10^{8}$ $L_\odot$) with high masses in the range $M = 20$ -- $80 M_\odot$. They undergo episodic outbursts with significant mass losses of about $10^{-3}$ -- $10^{-2}$ $M_\odot$ yr$^{-1}$~\citep{Taddia2013} for relatively high wind velocities ($u_\text{w} \sim 10^2$ km s$^{-1}$). Hence, the progenitor properties derived in this study might also indicate a LBV progenitor at the lower edge of the typical mass and luminosity spectrum, although this appears to be a less likely scenario.

\section{On the attenuation problem}
\label{sec:new} 
    At the time of writing, a comprehensive model for accurately defining the impact and timescale on which gamma-gamma interactions might attenuate the VHE flux to the point of being undetectable is missing. Current models adopt different approaches, often considering distinct parameter contributions, leading to various possible scenarios that strongly depend on the system's geometry and whether its evolution is treated as time-dependent or time-independent \citep[e.g.,][]{Cristofari2020}. In a recent work, \citet{Cristofari2022} modeled the gamma-ray emission of a typical II-P SN with results depending on the total explosion energy, the mass-loss rate of the pre-SN wind, the wind terminal velocity, the mass of the ejecta and the radius of the progenitor star. They show how the gamma-gamma attenuation significantly affects the flux above 100 GeV at early times (between 0 and 12 days), by potentially suppressing the expected VHE flux by about 20 orders of magnitude. However, the impact of this attenuation decreases with time, and at 12 days it can lower the effective flux by 1 to 5 orders of magnitude depending on the combination of the various parameters. Similarly, \citet{Brose2022} simulated the emission from II-P and IIn SNe associated with RSG and LBV progenitor stars, respectively, in the H.E.S.S. and \textit{Fermi}-LAT energy ranges. In both cases the gamma-gamma absorption can effectively suppress the flux at early times, with the effect being more pronounced at energies above 1 TeV. The gamma-ray luminosity can potentially be attenuated by 3 orders of magnitude at most at 15 days depending on the relative mass-loss rate. After the peak of luminosity, the effect can still be relevant up to $\sim$ 220 days.

    A significant factor in modeling gamma-gamma absorption is the evolution of the photosphere. As shown in Fig.~\ref{fig:phot}, the SN photosphere undergoes rapid cooling between the time of explosion and the bulk of the LST-1 observations, with its temperature decreasing by a factor of 2. Meanwhile, the radius expands almost linearly in the first days, with a sudden change of slope around day 15, possibly caused by a decoupling between the shock and the photospheric radius. This feature coincides with the onset of a decline in the pseudo-bolometric luminosity. These changes suggest that the optical target photon field becomes less dense during the bulk of LST-1 observations, supporting the approximation that absorbed and unabsorbed fluxes are approximately coincident in this phase. More broadly, the relatively rapid decline in optical luminosity observed in type II-L SNe may imply systematically weaker gamma-gamma absorption compared to the more extended plateau phase of type II-P SNe. However, a more quantitative description would require a dedicated model based on the observed evolution, which is beyond the scope of this work.
    
    Based on these considerations, it is possible that the gamma-flux observations of SN~2024bch cannot be directly converted to mass-loss ULs due to the role of the gamma-gamma attenuation. The LST-1 observations were performed starting at $t_0$ + 16, with the bulk of observations clustered around 20 days after the explosion, when the impact of gamma-gamma absorption may still have been relevant, attenuating the flux for a factor of three orders of magnitude at most. Considering these three levels of attenuation, our flux UL would translate to a weaker mass-loss-rate to wind-velocity ratio UL. In particular, assuming $F_{\text{abs}} = 10^{-x} F_{\text{unabs}}$, where $x = 1, 2, 3$, we obtain $\dot M/u_\text{w} \le [0.38, 1.2, 3.8] \times 10^{-3} \frac{\text{M}_\odot}{\text{yr}}\frac{\text{s}}{\text{km}}$, respectively.

\section{Conclusions} 
\label{sec:6}

    We have presented a multiwavelength analysis of SN~2024bch. We conducted a follow-up campaign at VHE gamma rays with the LST-1 and in the optical with the LT. No detection was achieved in the VHE band, and we computed the corresponding ULs for the six nights of observation with the LST-1. To interpret such ULs in a physical context, we considered the model from~\citet{Dwarkadas2013}, estimating ULs for the ratio of the mass-loss rate to the wind velocity. Moreover, we performed an optical analysis, employing one spectrum observed with the LT complemented with ancillary data from online public repositories for both photometry and spectroscopy. We used the open-source software \texttt{SuperBol} \citep{Nicholl2018} to study the photospheric evolution of SN~2024bch and the software \texttt{CASTOR} \citep{Simongini2024} to estimate the parameters of the event and to compare SN~2024bch light curves with those of other known events. Furthermore, we analyzed a pre-explosion image of SN~2024bch from the \textit{Hubble} Space Legacy Archive. Our major findings are as follows: 

    \begin{enumerate}[i.]
        \item The integrated photon flux UL of SN~2024bch above 100 GeV is $F_\gamma(> 100\, \mbox{GeV}) \le 3.61 \times 10^{-12}$ cm$^{-2}$ s$^{-1}$, as obtained from 12 h of LST-1 observations; this is coincident with a luminosity UL of $L_\gamma \le 1.33 \times 10^{41} \text{erg s}^{-1}$. These ULs align with those of other types of SNe found by \citet{Abdalla2019}, \citet{Ahnen2017}, and \citet{Acharyya2023}. However, none of the cited works included IIn-L SNe. Therefore, this is the first ever determined gamma-flux UL for a SN of this class, and the first ever observation of a CCSN by an IACT with such a low energy threshold.
        \item SN~2024bch was likely a type IIn-L SN according to several pieces of evidence. The light curves show a fast linear decay typical of type II-L SNe; the spectra evolve from IIn-like to IIL-like, with narrow and weak H$-\alpha$ profiles at early times that disappear after the peak of luminosity in favor of broad and strong lines at later times. According to \citet{Tartaglia2024}, the narrow features disappeared between day 16 and day 42 after the explosion. We find the photometric behavior to be similar to that of SN~2009kr, SN~2013fc, and SN~2008fq, while an independent classification made by \citet{Balcon2024}, performed around 3.5 days after the explosion, suggests a spectral similarity to SN~1998S.
        \item The SN~2024bch progenitor was likely a RSG according to our estimated parameters. From independent analyses we found $\text{M}_{\text{pr}} = (11$--$20)\, {M}_\odot$, $\text{R}_{\text{pr}} = 531 \pm 125 \, {R}_\odot$, $\text{L}_{\text{pr}} \le 10^{4.82}\,{L}_{\odot}$, $\text{T}_{\text{pr}} \le 4000\, \text{K,}$ and $\dot M/u_\text{w} \le  10^{-4} \frac{\text{M}_\odot}{\text{yr}}\frac{\text{s}}{\text{km}}$, all of which are in good agreement with the typical values of these stars. The RSG progenitor channel is also supported by the classification, as  type IIn-L SN explosion sites are found to be more similar to those of type II-L progenitors than those of type IIn \citep{Taddia2015}.
    \end{enumerate}
    
    Based on our interpretation, SN~2024bch belongs to the IIn-L SN class. Moreover, our results, in line with the existing literature, strongly suggest a RSG progenitor, although we cannot exclude the LBV progenitor channel. The uncertainty regarding the preferred progenitor channel for SN~2024bch highlights the crucial role of the mass-loss rate estimate. Future observations of SNe at VHEs will eventually reveal the elusive signal, enabling a more accurate modeling of the incoming flux and more constraining results.

\begin{acknowledgements}
We gratefully acknowledge financial support from the following agencies and organisations: Conselho Nacional de Desenvolvimento Cient\'{\i}fico e Tecnol\'{o}gico (CNPq), Funda\c{c}\~{a}o de Amparo \`{a} Pesquisa do Estado do Rio de Janeiro (FAPERJ), Funda\c{c}\~{a}o de Amparo \`{a} Pesquisa do Estado de S\~{a}o Paulo (FAPESP), Funda\c{c}\~{a}o de Apoio \`{a} Ci\^encia, Tecnologia e Inova\c{c}\~{a}o do Paran\'a - Funda\c{c}\~{a}o Arauc\'aria, Ministry of Science, Technology, Innovations and Communications (MCTIC), Brasil;
Ministry of Education and Science, National RI Roadmap Project DO1-153/28.08.2018, Bulgaria;
Croatian Science Foundation (HrZZ) Project IP-2022-10-4595, Rudjer Boskovic Institute, University of Osijek, University of Rijeka, University of Split, Faculty of Electrical Engineering, Mechanical Engineering and Naval Architecture, University of Zagreb, Faculty of Electrical Engineering and Computing, Croatia;
Ministry of Education, Youth and Sports, MEYS  LM2023047, EU/MEYS CZ.02.1.01/0.0/0.0/16\_013/0001403, CZ.02.1.01/0.0/0.0/18\_046/0016007, CZ.02.1.01/0.0/0.0/16\_019/0000754, CZ.02.01.01/00/22\_008/0004632 and CZ.02.01.01/00/23\_015/0008197 Czech Republic;
CNRS-IN2P3, the French Programme d’investissements d’avenir and the Enigmass Labex, 
This work has been done thanks to the facilities offered by the Univ. Savoie Mont Blanc - CNRS/IN2P3 MUST computing center, France;
Max Planck Society, German Bundesministerium f{\"u}r Bildung und Forschung (Verbundforschung / ErUM), Deutsche Forschungsgemeinschaft (SFBs 876 and 1491), Germany;
Istituto Nazionale di Astrofisica (INAF), Istituto Nazionale di Fisica Nucleare (INFN), Italian Ministry for University and Research (MUR), and the financial support from the European Union -- Next Generation EU under the project IR0000012 - CTA+ (CUP C53C22000430006), announcement N.3264 on 28/12/2021: ``Rafforzamento e creazione di IR nell’ambito del Piano Nazionale di Ripresa e Resilienza (PNRR)'';
ICRR, University of Tokyo, JSPS, MEXT, Japan;
JST SPRING - JPMJSP2108;
Narodowe Centrum Nauki, grant number 2019/34/E/ST9/00224, Poland;
The Spanish groups acknowledge the Spanish Ministry of Science and Innovation and the Spanish Research State Agency (AEI) through the government budget lines
PGE2022/28.06.000X.711.04,
28.06.000X.411.01 and 28.06.000X.711.04 of PGE 2023, 2024 and 2025,
and grants PID2019-104114RB-C31,  PID2019-107847RB-C44, PID2019-104114RB-C32, PID2019-105510GB-C31, PID2019-104114RB-C33, PID2019-107847RB-C43, PID2019-107847RB-C42, PID2019-107988GB-C22, PID2021-124581OB-I00, PID2021-125331NB-I00, PID2022-136828NB-C41, PID2022-137810NB-C22, PID2022-138172NB-C41, PID2022-138172NB-C42, PID2022-138172NB-C43, PID2022-139117NB-C41, PID2022-139117NB-C42, PID2022-139117NB-C43, PID2022-139117NB-C44, PID2022-136828NB-C42, PDC2023-145839-I00 funded by the Spanish MCIN/AEI/10.13039/501100011033 and “and by ERDF/EU and NextGenerationEU PRTR; the "Centro de Excelencia Severo Ochoa" program through grants no. CEX2019-000920-S, CEX2020-001007-S, CEX2021-001131-S; the "Unidad de Excelencia Mar\'ia de Maeztu" program through grants no. CEX2019-000918-M, CEX2020-001058-M; the "Ram\'on y Cajal" program through grants RYC2021-032991-I  funded by MICIN/AEI/10.13039/501100011033 and the European Union “NextGenerationEU”/PRTR and RYC2020-028639-I; the "Juan de la Cierva-Incorporaci\'on" program through grant no. IJC2019-040315-I and "Juan de la Cierva-formaci\'on"' through grant JDC2022-049705-I. They also acknowledge the "Atracci\'on de Talento" program of Comunidad de Madrid through grant no. 2019-T2/TIC-12900; the project "Tecnolog\'ias avanzadas para la exploraci\'on del universo y sus componentes" (PR47/21 TAU), funded by Comunidad de Madrid, by the Recovery, Transformation and Resilience Plan from the Spanish State, and by NextGenerationEU from the European Union through the Recovery and Resilience Facility; “MAD4SPACE: Desarrollo de tecnolog\'ias habilitadoras para estudios del espacio en la Comunidad de Madrid" (TEC-2024/TEC-182) project funded by Comunidad de Madrid; the La Caixa Banking Foundation, grant no. LCF/BQ/PI21/11830030; Junta de Andaluc\'ia under Plan Complementario de I+D+I (Ref. AST22\_0001) and Plan Andaluz de Investigaci\'on, Desarrollo e Innovaci\'on as research group FQM-322; Project ref. AST22\_00001\_9 with funding from NextGenerationEU funds; the “Ministerio de Ciencia, Innovaci\'on y Universidades”  and its “Plan de Recuperaci\'on, Transformaci\'on y Resiliencia”; “Consejer\'ia de Universidad, Investigaci\'on e Innovaci\'on” of the regional government of Andaluc\'ia and “Consejo Superior de Investigaciones Cient\'ificas”, Grant CNS2023-144504 funded by MICIU/AEI/10.13039/501100011033 and by the European Union NextGenerationEU/PRTR,  the European Union's Recovery and Resilience Facility-Next Generation, in the framework of the General Invitation of the Spanish Government’s public business entity Red.es to participate in talent attraction and retention programmes within Investment 4 of Component 19 of the Recovery, Transformation and Resilience Plan; Junta de Andaluc\'{\i}a under Plan Complementario de I+D+I (Ref. AST22\_00001), Plan Andaluz de Investigaci\'on, Desarrollo e Innovación (Ref. FQM-322). ``Programa Operativo de Crecimiento Inteligente" FEDER 2014-2020 (Ref.~ESFRI-2017-IAC-12), Ministerio de Ciencia e Innovaci\'on, 15\% co-financed by Consejer\'ia de Econom\'ia, Industria, Comercio y Conocimiento del Gobierno de Canarias; the "CERCA" program and the grants 2021SGR00426 and 2021SGR00679, all funded by the Generalitat de Catalunya; and the European Union's NextGenerationEU (PRTR-C17.I1). This research used the computing and storage resources provided by the Port d’Informaci\'o Cient\'ifica (PIC) data center.
State Secretariat for Education, Research and Innovation (SERI) and Swiss National Science Foundation (SNSF), Switzerland;
The research leading to these results has received funding from the European Union's Seventh Framework Programme (FP7/2007-2013) under grant agreements No~262053 and No~317446;
This project is receiving funding from the European Union's Horizon 2020 research and innovation programs under agreement No~676134;
ESCAPE - The European Science Cluster of Astronomy \& Particle Physics ESFRI Research Infrastructures has received funding from the European Union’s Horizon 2020 research and innovation programme under Grant Agreement no. 824064. 
\\
We acknowledge with thanks the variable star observations from the AAVSO International Database contributed by observers worldwide and used in this research. Based on observations made with the NASA/ESA \textit{Hubble} Space Telescope, and obtained from the \textit{Hubble} Legacy Archive, which is a collaboration between the Space Telescope Science Institute (STScI/NASA), the Space Telescope European Coordinating Facility (ST-ECF/ESA) and the Canadian Astronomy Data Centre (CADC/NRC/CSA). This article is also based on observations made in the Liverpool telescope located at the Observatorio del Roque de los Muchachos under the CL24A06 (PI: A. L\'opez-Oramas) program. This research is part of the Project RYC2021-032991-I, funded by MICIN/AEI/10.13039/501100011033, and the European Union “NextGenerationEU”/PRTR. 
\\ \\ 
\textit{Author contribution:} A. Simongini: project coordination, LST-1 data analysis, optical data analysis, model fitting, physical interpretation, paper drafting and edition; A. López-Oramas: PI of the LST-1 proposal, PI of the LT proposal, discussion of the model and of the obtained results, paper edition; A. Carosi: discussion of the model and of the obtained results, paper edition; A. Aguasca-Cabot: LST-1 data analysis, paper edition. The rest of the authors have contributed in one or several of the following ways: design, construction, maintenance and operation of the instrument(s) used to acquire the data; preparation and/or evaluation of the observation proposals; data acquisition, processing, calibration and/or reduction; production of analysis tools and/or related Monte Carlo simulations; discussion and approval of the contents of the draft.

\end{acknowledgements}

\bibliographystyle{aa}
\bibliography{Biblios}

\end{document}